\documentclass[prl,twocolumn,superscriptaddress,%
preprintnumbers,amsmath,amssymb]{revtex4}
\usepackage{amsmath}
\usepackage{graphicx}
\usepackage{dcolumn}

\begin{document}

\title{Interferometry of Klein tunnelling electrons in graphene quantum rings}

\author{D. J. P. de Sousa}\email{duarte@fisica.ufc.br}
\affiliation{Departamento de F\'isica, Universidade
Federal do Cear\'a, Caixa Postal 6030, Campus do Pici, 60455-900
Fortaleza, Cear\'a, Brazil}
\author{Andrey Chaves}\email{andrey@fisica.ufc.br}
\affiliation{Departamento de F\'isica, Universidade
Federal do Cear\'a, Caixa Postal 6030, Campus do Pici, 60455-900
Fortaleza, Cear\'a, Brazil}
\affiliation{Department of Chemistry, Columbia University, 3000 Broadway, New York, New York 10027, USA}
\author{J. M. Pereira Jr.}\email{pereira@fisica.ufc.br}
\author{G. A. Farias}\email{gil@fisica.ufc.br}
\affiliation{Departamento de F\'isica, Universidade
Federal do Cear\'a, Caixa Postal 6030, Campus do Pici, 60455-900
Fortaleza, Cear\'a, Brazil}

\date{ \today }

\begin{abstract}
We theoretically study a current switch that exploits the phase acquired by a charge carrier as it tunnels through a potential barrier in graphene. The system acts as an interferometer based on an armchair graphene quantum ring, where the phase difference between interfering electronic wave functions for each path can be controlled by tuning either the height or the width of a potential barrier in the ring arms. By varying the parameters of the potential barriers the interference can become completely destructive. We demonstrate how this interference effect can be used for developing a simple graphene-based logic gate with high \textit{on/off} ratio.

\end{abstract}

\maketitle

\section{Introduction}
Monolayer graphene, a one atom thick carbon-based material with high electronic mobility \cite{CastroNetoReview} has been considered a promising candidate to replace Silicon in future nanodevices since its first fabrication in 2004 \cite{Geim}. However, the chiral nature of the charge carriers in graphene, together with their gapless spectrum, gives rise to the phenomenon of Klein tunneling, i.e. the perfect transmission of normally-incident electrons through a potential barrier \cite{Katsnelson, Nosso}. This unusual phenomenon causes difficulties for the creation of logic devices, due to the fact that, for wide graphene samples it may not be possible to obtain an "off" state, in which there is no flow of current. The ability to shut down the charge transport throughout the device in a controllable way, which is easily achieved in semiconductor junctions, is fundamental for the development of future graphene-based electronics. In this sense, most of the efforts to design graphene logic devices nowadays aim towards producing a gap in the system, e.g. by introducing a bias in bilayer graphene \cite{CastroNetoReview} or by strain or edge engineering in monolayer graphene \cite{Geim, Katsnelson, Guinea}. Other proposals have considered using gate voltages around the charge neutrality point to realize logic operations \cite{Russo} and biased nanoribbon crossings to achieve current switching \cite{Lake}. 

In this work we demonstrate the possibility of obtaining controllable current switching with high \textit{on/off} ratio in a monolayer graphene ring without the need of a gap in its spectrum, by exploiting interference effects in a quantum ring in the ballistic regime. Graphene-based quantum rings have been previously investigated, both theoretically and experimentally (see, e.g. Ref. \onlinecite{Trauzettel, Daniela, Brey, Cabosart}) and have been shown to display unusual features in their magnetoconductance, due to the massless Dirac Fermion character of the charge carriers in this system. The feasibility of mesoscopic devices based on phase-coherent effects is grounded on the fact that graphene structures can display large phase relaxation lengths \cite{PhaseRelaxationLength} and previous work has suggested the possibility of using side gates to control the relative phase of electrons transported through each arm of a graphene ring in order to obtain a transistor effect \cite{Adame}.

The idea of harnessing quantum interference effects to control the current flow of a nanodevice has also been previously raised in the context of molecular electronics. Stafford \textit{et al}. \cite{Mazumdar} considered a system consisting of a monocyclic aromatic molecule connected to external leads. The current would be controlled by the introduction of decoherence due to the proximity of a scanning transmission microscope tip. Saha et al. \cite{Saha} proposed a molecular transistor where graphene nanoribbons act as electrodes connected to a ring shapped 18-annulene molecule. However, the working of these devices would depend on the precise positioning of the leads on particular molecular sites, which may not be suitable for practical technological applications.

\section{Model}
\begin{figure}[!bpht]
\centerline{\includegraphics[scale = 0.4]{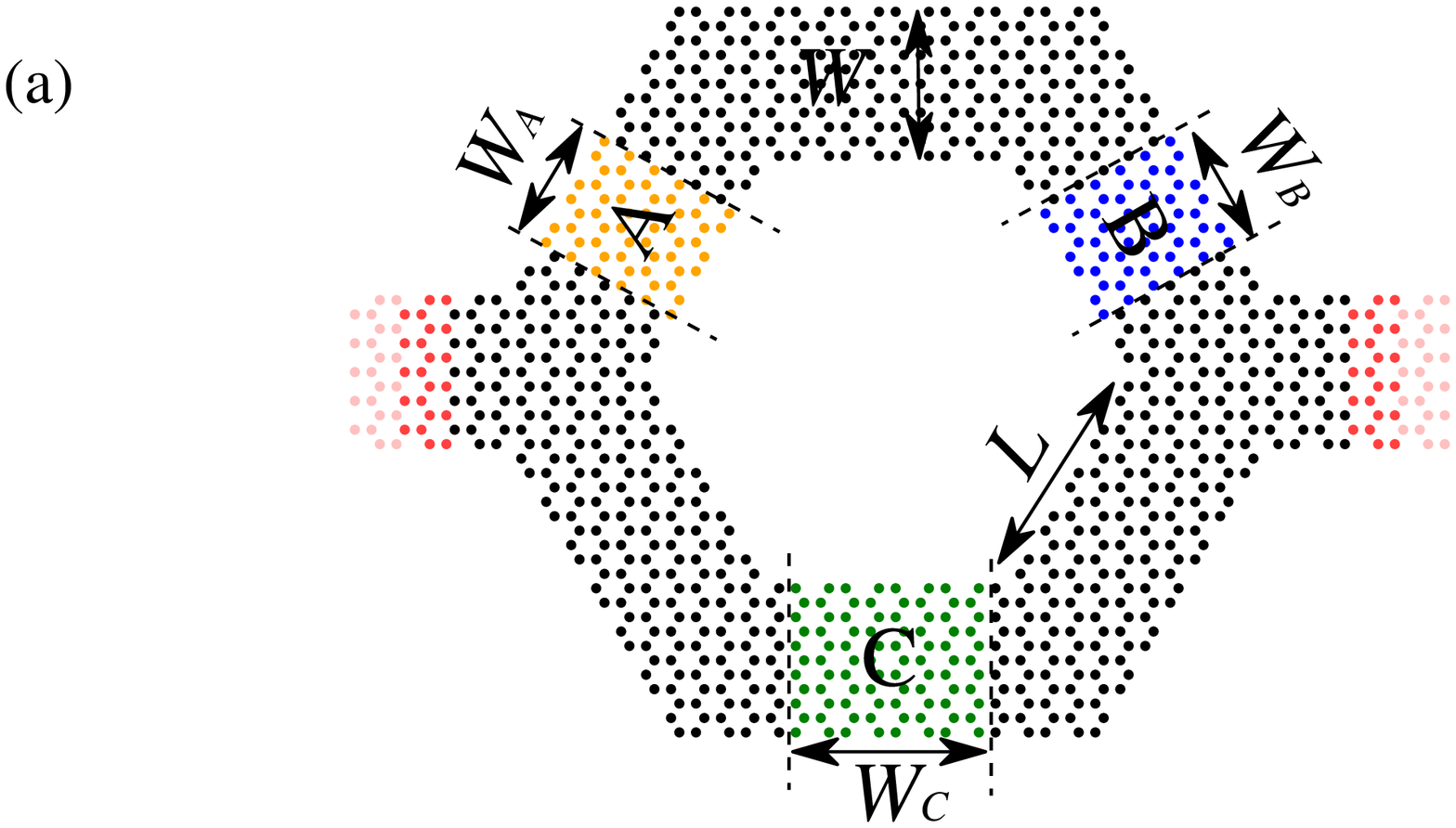}}
\centerline{\includegraphics[width = \linewidth]{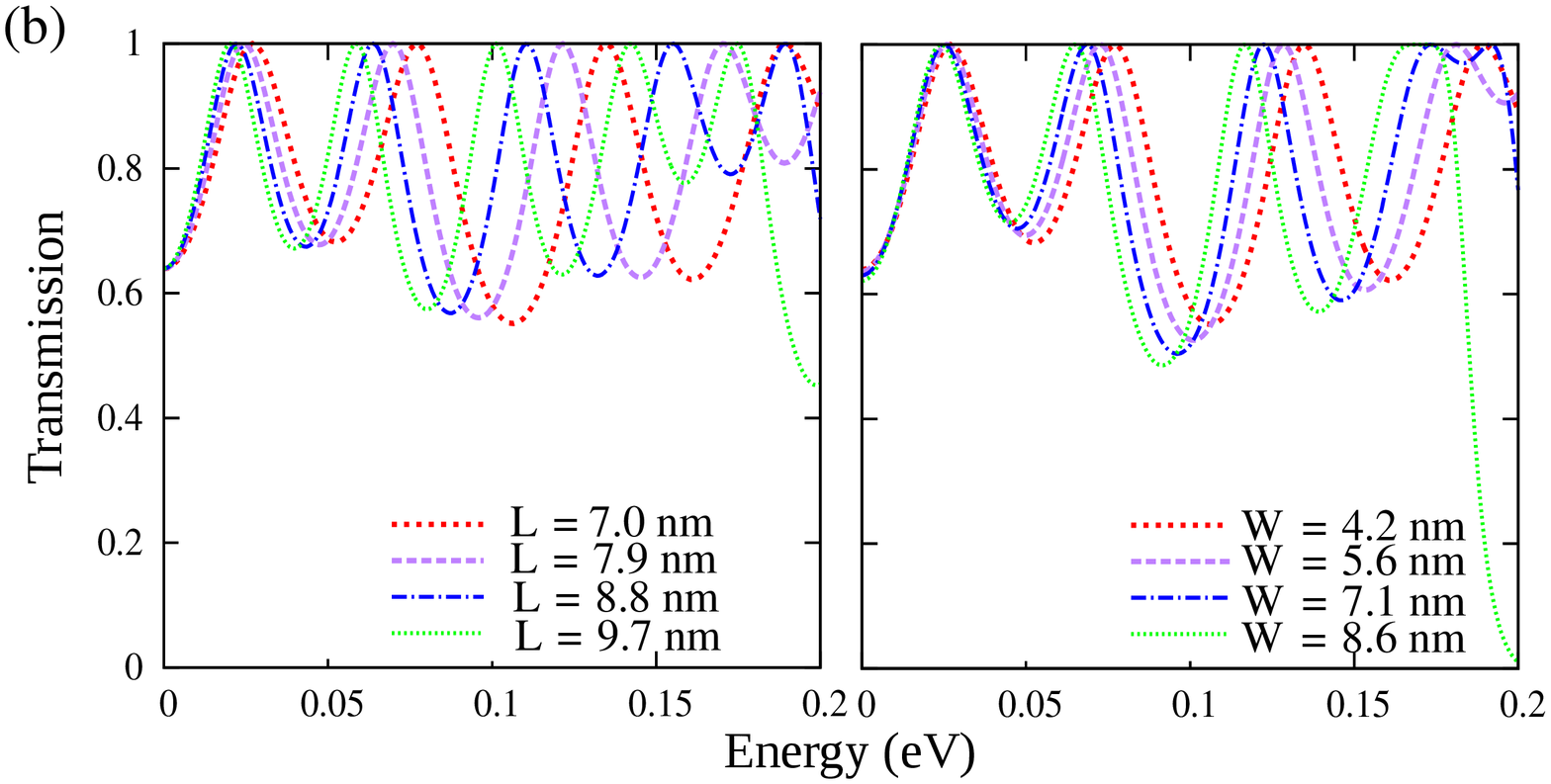}}
\caption{(Color online) (a) Sketch of the proposed interferometer: an hexagonal armchair ring, attached to infinite input and output leads (in fading red), with three regions of non-zero potential, produced by electrodes, labelled as A and B, in the upper arm, and C, in the lower arm of the ring. (b) Transmission probabilities as a function of energy in such an interferometer, considering $V_A = V_B = V_C = 0$, for different values of arms length $L$ with a fixed width $W = 5.96 $ nm (left), and for different values of width and fixed length $L = 9.94 $ nm (right).} \label{fig:Fig1}
\end{figure}

The system investigated here consists of a mesoscopic carbon hexagonal ring, (see Fig. \ref{fig:Fig1}(a)), made of \textit{metallic} armchair graphene nanoribbons. 
It has been recently shown that the electronic states of such hexagonal armchair rings closely resemble those of massless Dirac Fermions in an ideal circular 1D ring \cite{daCosta}. That indicates that edge effects may not be relevant for the description of charge transport in these systems, and that phase coherence effects can play an important role. In contrast with the system described in Ref. \onlinecite{Adame}, the present setup allows for the control of the electronic phase for each arm of the ring independently. For that, step potential barriers created by gate-induced \textit{pnp} junctions, labelled as A, B and C in Fig. \ref{fig:Fig1}(a), with widths $W_A$, $W_B$ and $W_C$, respectively, are placed in the ring arms. We assume that such thin gates over the ring arms do not affect the states coherence. After tunnelling through a potential barrier with high probability, due to the Klein tunnelling effect for low energy electrons in graphene, the electron wave function acquires a phase 
\begin{equation} \label{Eq.phase}
\phi = \frac{U_iW_i}{\hbar v_F},
\end{equation}
for $i$ = A, B or C, which depends on both the potential height $U_i$ and width $W_i$ of the barrier. \cite{Russo, Yacoby} This is the main mechanism behind the current control provided by the system considered here, as shown below. 


\section{Results}
Let us first investigate the transmission probabilities through such an interferometer as function of the electron energy in the absence of any external potential. Throughout this paper, the transmission probabilities $\mathcal{T}$ are calculated using a mode matching technique, by means of the software package KWANT \cite{kwant} which are then used for obtaining conduction by Landauer's formula. Results in Fig. \ref{fig:Fig1}(b) show this quantity as function of energy for a ring with arms widths $W = 5.96$ nm (left) and different values of arms length $L$, as well as for a fixed length $L = 9.94$ nm and several values of width $W$ (right). In all cases, one observes a non-zero transmission even for $E = 0$, which emphasizes the metallic character of the system. Quantum resonances through the system are found, as evident by the $\mathcal{T} = 1$ peaks for several values of electron energy. The energies where resonant peaks occur decrease by increasing either the length of the ring arms or their widths, labelled respectively as $L$ and $W$ in Fig. \ref{fig:Fig1}(a). The peaks are approximately equally spaced, which is reminiscent of the energy spectra of hexagonal armchair rings, whose energy levels can be predicted to a good approximation by the analytical expression $E_n = (\hbar v_F / R )(n + 1/2)$ with $n = 0, 1, 2, ...$, where $R$ is the radius of a circle with the same area as the hexagon whose sides are given by the average length between the inner and outer edges of the hexagonal ring \cite{daCosta}. Indeed, for a ring with dimensions $L = 44.4 $ nm and $W = 18.3 $ nm, larger than those in Fig.~\ref{fig:Fig1}, such analytical expression yields $E_1$ = 0.0072 eV, $E_2$ = 0.0218 eV, and $E_3$ = 0.0364 eV, close to the numerically obtained values of maximum transmission (0.00625, 0.0175 and 0.0305 eV, respectively). 

\begin{figure}[!bpht]
\centerline{\includegraphics[width = \linewidth]{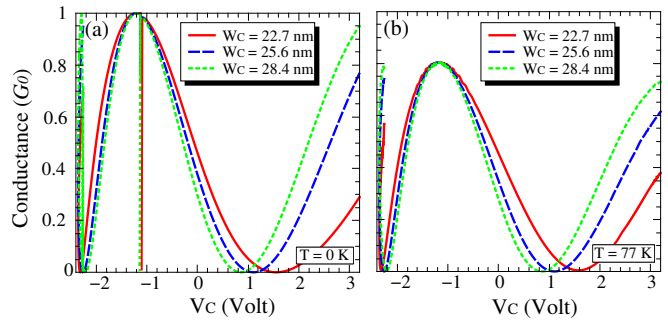}}
\caption{(Color online) Conductance in units of $G_{0} = 2e^{2}/h$ as a function of the gate voltage $V_{C}$ in region C [see Fig. 1(a)], with $W = 14.2$ nm and $L = 38.9$ nm, for $V_A = V_B = 0$, considering different barrier width $W_C$ for (a) T = 0 K and (b) T = 77 K.} \label{fig:Fig2}
\end{figure}

\begin{figure}[!bpht]
\centerline{\includegraphics[width = \linewidth]{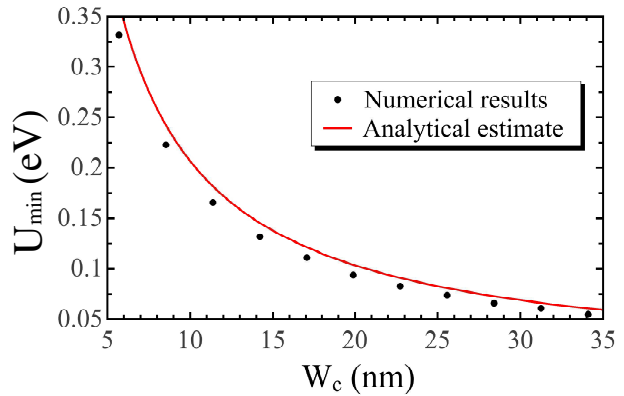}}
\caption{(Color online) Comparison between the numerically obtained (symbols) potential height $U_{min}$, where the transmission probability reaches 0, and its analytical estimate (solid curve).} \label{fig:Fig3}
\end{figure}

We now investigate the first resonance of a system with $L = 38.9 $ nm and $W = 14.2 $ nm, namely $E = 0.0068$ eV, in the presence of potential barriers. Henceforth, these will be the values of length and width considered in all of our results. These values were chosen according to the rule $L + 2W/\sqrt{3} \approx \ell_{e}/2$, where $\ell_{e} = 0.1 \ \ \mu$m is the mean free path of graphene at room temperature \cite{Geim}. The condition ensures ballistic transport properties for electrons in the central region of the system. In order to make our calculations more realistic, we assume that the extremities of the potential barriers decay smoothly, as a sinusoidal function, with a decay length $l_{decay} \approx 1.42$ nm. We have also included a realistic gate voltage $V_{i}$ necessary to tune the potential barrier height in region $i$ to the theoretical value $U_{i}$ ($i =$A, B or C). These calculations were based on experimental values of capacitance per area from Hall effect measurements, assuming a density of charge carriers on the back gate region of $n_{bg} \approx 10^{12}$ cm$^{-2}$, which is the accessed charge carrier density on a typical $\mathrm{Si}/300$ nm $\mathrm{SiO_{2}}$ substrate\cite{Scalable}. The functional relation is $V_{i}= (e/C_{bg})[(u_{i}  + \sqrt{n_{bg}})^{2} - 2n_{bg}]$, where $C_{bg} = 13.6$ nF$\cdot$cm$^{-2}$ is the mentioned back gate capacitance per area, $e$ is the electron charge and $u_{i} = U_{i}/(\sqrt{\pi}\hbar v_{F})$\cite{Stander}. We first take $V_A = V_B = 0$ and vary $V_C$, in order to tune the phase of the electrons travelling through the lower arm of the ring structure. The calculated conductance for this situation is shown as function of $V_C$ in Fig. \ref{fig:Fig2}(a), with $W_C = 22.7$ (solid), 25.6 (dashed) and 28.4 nm (dotted). The transmission exhibits a sharp minimum for $T = 0$ K around $U_C = E$ for all cases. This is due to the fact that Klein tunnelling probability approaches zero faster as the electron deviates from the normal incidence when the potential barrier height is close to the electron energy \cite{Greiner}. Such suppression of the transmission is not controllable by external parameters and is negligible for non-zero temperature, as one can verify in Fig. 2(b). In fact, for higher temperatures, a broader range of energies contribute to the conductance and most of the energies are out of this resonance, which is a very sharp peak. Conversely, there is a clear oscillating background in each curve, leading to zero transmission points for specific values of potential height $U_{min}$, which are controllable by the barrier width $W_C$, as shown by the different curves in Fig. \ref{fig:Fig2}. These minima are preserved even for higher temperatures, as shown in Fig. 2(b). We argue that such a transmission minimum is a consequence of the phase in Eq. (\ref{Eq.phase}) acquired by the electron that tunnels through the potential barrier. This occurs when
\begin{equation} \label{Eq.phipi}
U_{min} =\frac{\pi\hbar v_F}{W_C},
\end{equation} 
A comparison between the numerically obtained first minimum $U_{min}$ (symbols) and those predicted by the analytical expression (curve) in Eq.(\ref{Eq.phipi}) is shown in Fig. \ref{fig:Fig3} as function of the barrier width in the lower arm. Fairly good agreement between these results lends support to the idea that Klein tunnelling phase is indeed behind the current modulation observed in numerically obtained results. The agreement improves as $W_C$ increases, therefore, in order to be able to predict $U_{min}$, and thus $V_{min}$, to a good precision with the analytical expression in Eq. (\ref{Eq.phipi}) one would need long ring arms, which can accommodate larger potential barrier widths.

This result demonstrates the first possibility of application of the interferometer in Fig. \ref{fig:Fig1}(a): a OR logic gate. For this, we set $W_A = W_B = W_C /2$ and select $U_{C} = U_{min}$. In this case, since the induced phase is proportional to the barrier width, as barrier C produces a $\pi$ phase for the same $U_{min}$, barriers A and B would produce a $\pi/2$ phase each. Initially, with $U_A = U_B = 0$, there is no current, since the potential barrier in the lower arm is such that the interference is completely destructive at the right lead. Now, if $U_{A} = 2U_{min}$ and $U_{B} = 0$, the relative phase difference between electrons travelling through the different paths is zero and the transmission probability reaches a maximum 
That can also be obtained if $U_{B} = 2U_{min}$ and $U_{A} = 0$, or $U_A = U_B = U_{min}$. In short, if $U_{A}$ \textit{or} $U_{B}$, \textit{or} both, are selected, the currents flows. As previously mentioned, one can also control the critical potential $U_{min}$ for shutting down the current by tuning the potential barrier width. 

\begin{figure}[!bpht]
\centerline{\includegraphics[width = \linewidth]{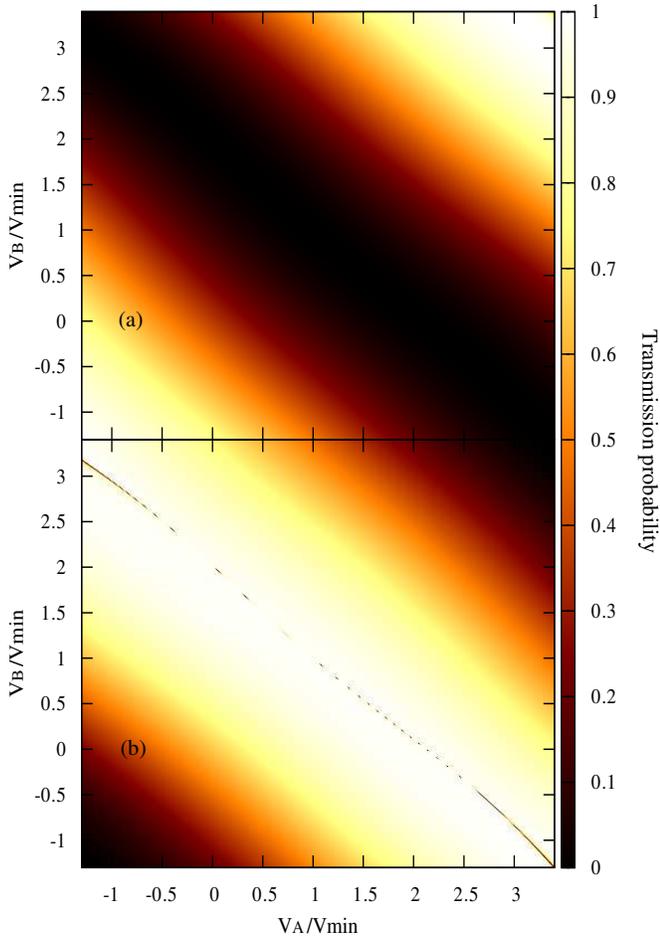}}
\caption{(Color online) Contour plots of the transmission probability as a function of the gate voltage $V_A$ and $V_B$ in regions A and B [see Fig. 1(a)], considering barrier widths $W_A = W_B = 14.2$ nm and $ W_C = 28.4$ nm, for two values of gate voltage in C: $V_C = 0$ (a) and $V_C = V_{min} = 0.862$ V (b).} \label{fig:Fig4}
\end{figure}

The situation is illustrated in Fig. \ref{fig:Fig4} by the contour plots of $\mathcal{T}$ as function of $V_A/V_{min}$ and $V_B/V_{min}$, for  $W_A = W_B = 14.2$ nm and $W_C = 28.4$ nm. For a 28.4 nm barrier width, one obtains $U_{min} \approx 0.076$ eV from Eq. (\ref{Eq.phipi}), by considering $W_{C} \rightarrow W_{C} - l_{decay}$, which corresponds to a gate voltage $V_{min} = 0.862$ V. In Fig. \ref{fig:Fig4}(b), for $V_C = 0.862$ V, a region of strongly reduced transmission probability ($\mathcal{T} \approx$ 0, darker colors) is seen around $V_A = V_B = -e n_{bg}/C_{bg}$ (corresponding to $U_A = U_B = 0$), whereas higher values of $\mathcal{T}$ are found if only one of the potential barriers is increased, while the other is kept at zero. As $U^{(1)}_{min}$, which is the height of the potential barrier for the first minimum, is inversely proportional to the barrier width, and $W_A$ and $W_B$ are each half the width of $W_C = 28.4$ nm, the system can be set so that the electrodes alone are able to provide full current only when a single electrode (A or B) is set at $\approx 2U^{(1)}_{min} = 0.152$ eV, corresponding to a gate voltage of $3.8$ V, as one can see by the bright regions in the color plot in Fig. \ref{fig:Fig4}(b) around $V_{A(B)} = 3.8$ V and $V_{B(A)} = 0$ V. Such high current is also obtained when both electrodes are at 0.862 V ($V_{A(B)}/V_{min} = 1$). Notice that current is suppressed again if both electrodes are at 3.8 V instead, which opens the way to produce a XOR logic device, where current is non-zero only if electrodes A OR B are turned on, but not when both are on. Similar results are easily obtained if one considers $V_C = 0$ and tunes $V_A$ and $V_B$ accordingly, as shown in Fig. \ref{fig:Fig4}(a). 

It is also possible to use the proposed interferometer as a AND gate: for this, let us consider $U_C = U_{min}$. As before, there is no current if $U_A = U_B = 0$. Now, using $U^{(2)}_{min} (\approx 2U_{min})$ as height of the potential barrier in both electrodes A and B (simultaneously) for which there is a second minimum of transmission when $U_{C} = U_{min}$, current is suppressed by selecting $U_A = 2U^{(2)}_{min}$ and $U_B = 0$, or $U_B = 2U^{(2)}_{min}$ and $U_A = 0$. The only way a maximum of transmission can be found in this case is when $U_{A} = U_{min}$ AND $U_{B} = U_{min} $ at the same time, hence, providing a AND gate.

Table I summarizes all possible logic gates by defining $\alpha = U_{min}^{(2)}/U_{min}$, $\beta = U_{max}/U_{min}$, $u_{A} = U_{A}/U_{min}$ and $u_{B} = U_{B}/U_{min}$, where $U_{min}^{c}$ and $U_{max}$ are the values of $U_{A} (U_{B})$ for which we have the second minimum of transmission when $U_{C}=U_{min}$ and the second maximum when $U_{C} = 0$, respectively.

\begin{table}[h]
\begin{tabular}{|ccc|ccc|ccc|}
\hline
\multicolumn{3}{|c|}{AND ($U_{C}=U_{min}$)}  & \multicolumn{3}{|c|}{NAND ($U_{C}=0$)}  & \multicolumn{3}{|c|}{OR ($U_{C}=U_{min}$)}   \\
\hline \hline

 \ \ $u_{A}$ \ \ &  \ \ $u_{B}$ \ \  &  \ \ $T$  \ \  & \ \ $u_{A}$ \ \ & $u_{B}$ \ \  &  \ \ $T$  \ \   & \ \ $u_{A}$ \ \ &  \ \ $vu_{B}$ \ \  & \ \  $T$ \ \ \\ \hline 
\ \ 0 \ \ &\ \ 0 \ \ & \ \ 0 \ \  & \ \ 0 \ \ & \ \ 0 \ \ & \ \ 1  \ \  & \ \ 0 \ \ & \ \ 0 \ \ & \ \ 0  \ \ \\ \hline
 \ \ 2$\alpha$ \ \  & \ \  0 \ \  &  \ \ 0 \ \   &  \ \ $2\beta$ \ \  &  \ \ 0  \ \ &  \ \ 1  \ \  & \ \  2 \ \  & \ \  0 \ \  & \ \  1 \ \   \\ \hline
 \ \ 0  \ \ &  \ \ 2$\alpha$ \ \  &  \ \ 0  \ \  &  \ \ 0  \ \ &  \ \ $2\beta$ \ \  & \ \  1 \ \   & \ \  0 \ \  & \ \  2 \ \  & \ \  1 \ \   \\ \hline
 \ \ 1 \ \  & \ \  1 \ \  & \ \  1 \ \   & \ \  1 \ \  & \ \  1 \ \  & \ \  0 \ \   & \ \  1 \ \  & \ \  1 \ \  & \ \  1 \ \   \\ \hline \hline
 
 \multicolumn{3}{|c|}{NOR ($U_{C}=0$)}  & \multicolumn{3}{|c|}{XOR ($U_{C}=U_{min}$)}  & \multicolumn{3}{|c|}{XNOR ($U_{C}=0$)}   \\
\hline \hline
 \ \ $u_{A}$ \ \ &  \ \ $u_{B}$  \ \ & \ \  $T$ \ \   & \ \ $u_{A}$ \ \ &  \ \ $u_{B}$ \ \  &  \ \ $T$ \ \   & \ \ $u_{A}$ \ \ &  \ \ $u_{B}$  \ \ &  \ \ $T$ \ \  \\ \hline 
 \ \ 0 \ \  & \ \  0 \ \  & \ \  1  \ \   & \ \  0 \ \  & \ \  0 \ \  & \ \  0  \ \  & \ \  0 \ \  & \ \  0 \ \  & \ \  1 \ \   \\ \hline
 \ \ 2  \ \ &  \ \ 0 \ \  &  \ \ 0  \ \  &  2 &  \ \ 0  \ \ &  \ \ 1 \ \   & \ \  2 \ \  & \ \  0 \ \  & \ \  0 \ \   \\ \hline
 \ \ 0  \ \ &  \ \ 2  \ \ & \ \  0  \ \  & \ \  0 \ \  & \ \  2 \ \  & \ \  1 \ \   & \ \  0 \ \  & \ \  2 \ \  & 0  \\ \hline
 \ \ 1 \ \  &  \ \ 1  \ \ &  \ \ 0  \ \  &  \ \ $\alpha$ \ \  & \ \  $\alpha$ \ \  & \ \  0 \ \   &  \ \ $\beta$  \ \ & \ \  $\beta$ \ \  & \ \  1 \ \   \\ \hline
\end{tabular}
\caption{Configuration of the potentials at the gates A, B and C [see Fig. \ref{fig:Fig1}(a)] for all cases of logic gates that are possible to be obtained with such a structure.}
\end{table}

Transmission probabilities in Fig.~\ref{fig:Fig4} are directly related to the conductance only in the case of zero temperature. At $T = 77$ K, the linear response formula for the conductance is used to obtain the results shown in Fig. \ref{fig:Fig2}(b). 
The first minimum of transmission predicted by the Klein tunneling phase shift, Eq. (\ref{Eq.phipi}), remains unaltered. Therefore, although the maximum value of the conductance equals $\approx 0.8 (2e^2 /h)$ for $T = 77$ K [see Fig.~\ref{fig:Fig2}(b)], the {\it on/off} ratio is still of the order of $10^{5}$, whereas the seconder order minimum exhibits {\it on/off} ratio with 4 orders of magnitude. Following Table I, four of the six logic gates presented, namely OR , NOR, NAND e XNOR, remain almost "intact" for non-zero temperatures, since they do not depend on the second order minimum. In a real device, the \textit{on/off} ratio will likely be affected by the presence of disorder, such as lattice defects or edge roughness. However, since the dimensions of the present system are smaller then the typical phase coherence length of electrons in graphene \cite{PhaseRelaxationLength}, one can expect the values of the \textit{on/off} ratio approaching the ones above may be measured.

At room temperature $T \approx 300$ K, one has an energy distribution around the Fermi level $E_F$, whose width is around $k_BT \approx 0.026$ eV. This means that, although for specific configurations of Fermi energy and potential the transmission is zero, the conductance will still be finite in these cases, due to contributions of electrons with energy around $E_F$. Nevertheless, the transmission probability minima in Fig. \ref{fig:Fig2} (a) are very smooth, so that in the vicinity of the minima, the probability is still close to zero, thus the transmission probability integrated in energy would still vanish and the \textit{on/off} ratio of the interferometer would remain high.

\begin{figure}[!bpht]
\centerline{\includegraphics[width = \linewidth]{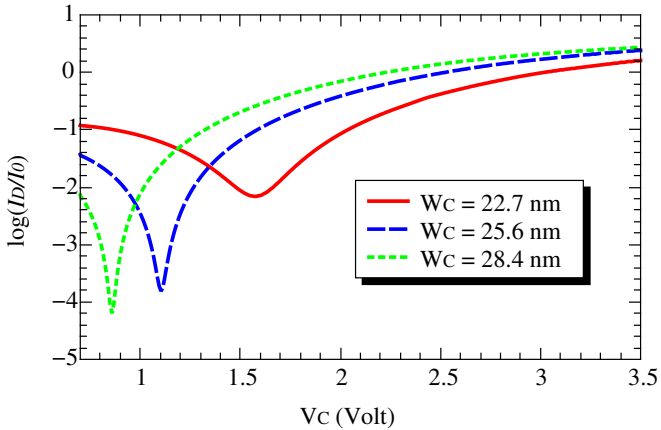}}
\caption{(Color online) Logarithm of the drain current $I_{D}$ in units of $I_{0} = 2e^{2}/h$ A as a function of the gate voltage $V_{C}$ in region C, with $W = 14.2$ nm and $L = 38.9 $ nm, for $V_{A} = V_{B} = 0$, considering different barrier width $W_{C}$.} \label{fig:Fig5}
\end{figure}

To estimate how fast the system leaves the sub-threshold region we computed the logarithm of the drain current as a function of gate voltage for different barrier widths at $T = 77$ K. The results in Fig.~\ref{fig:Fig5} show a clear dependence of the slope of the approximate linear region of the curves with the potential barrier widths, indicating that the sub-threshold swing becomes smaller with the increase of $W_{C}$. These calculation were based on the linear response formula limit, which allows us to write $\log(I/I_0) = \log(G) + \log(G_0 V_C / I_0)$, where $G = \int dE (-\partial f / \partial E) \mathcal{T}(E)$ with $f$ being the Fermi-Dirac distribution, and $I_0 = G_0 \times 1$ Volt. 

\begin{figure}[!bpht]
\centerline{\includegraphics[width = \linewidth]{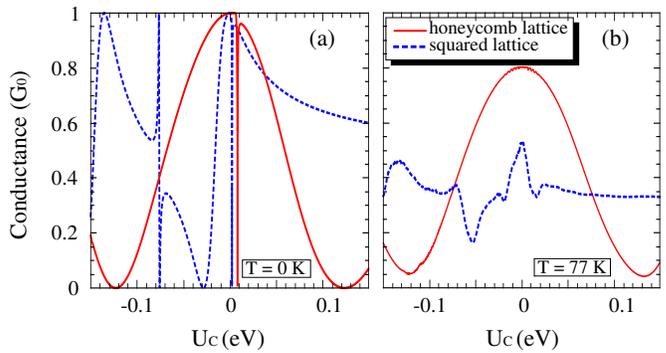}}
\caption{(Color online) A comparison between the conductance for a hexagonal quantum ring for two different lattices: squared lattice (dashed blue curve) and honeycomb lattice (solid red curve) at (a) T = 0 K and (b) T = 77 K.  } \label{fig:Fig6}
\end{figure}

It is important to emphasize that the oscillating behaviour of the conductance for positive and negative values of the potential barrier height in region C is a consequence of the unique behaviour of electrons in graphene and it is not expected to be present in materials with parabolic dispersion. Indeed, the transmission of a schr\"odinger electron through a potential barrier decreases monotonically by increasing the potential height, $U$, assuming that the initial electron energy, $E$, obeys $E < U$. For $E > U$, the transmission oscillates reaching a maximum whenever the condition for the resonant tunnelling is satisfied. In order to compare such characteristics with the graphene case, we plotted the conductance as a function of the potential barrier in region $C$ for a hexagonal quantum ring made out of a square lattice. The dimensions of the system are approximately the same we have used in the graphene case. In Fig.~\ref{fig:Fig6}(a), the conductance as function of $U_C$ for both, the square lattice case (dashed blue curve) and the honeycomb lattice case (solid red curve), at $T = 0 K$ is presented. The energy of the incident electron in the square lattice is chosen to be $E =  0.0205$ eV, which corresponds to a maximum of transmission when $U_C = 0$ eV for this system. For $U_C > E$, the transmission decreases monotonically and for $U_C < E$ an oscillating behaviour is observed, as one would expect, whereas the transmission oscillations for the honeycomb lattices is symmetric in relation to the $U_C = 0$ eV point. The energy range out of which the maximum and minimum of transmission occur for the square lattice case are very narrow, making such characteristics not appropriate for current modulation in such systems, since the effect of temperature is to smooth the conductance peaks, as shown in Fig.~\ref{fig:Fig6}(b).

\section{Conclusions}
In summary, we have calculated the transport properties of an armchair graphene quantum ring with a potential barrier in one of the ring arms. As the electron tunnels through the barrier due to Klein effect, it acquires a phase shift that is proportional to both the barrier height and width. As we assume the latter to be fixed, we demonstrate that the former can be adjusted so that we obtain a $\pi$ phase difference between electrons travelling through different arms of the ring, leading to full destructive interference and, therefore, zero current in the outgoing lead of the structure. Such an easily controllable interference effect provides the possibility of re-designing the system investigated here as to produce AND, NAND, OR, NOR, XOR and XNOR logic gates with monolayer graphene rings, which has fundamental importance for the development of future graphene-based electronics.  In other words, the Klein tunnelling effect, which has always been seen as a difficulty to be overcome in future graphene-based electronics, is actually a fundamental principle of operation for the interferometer studied here. Despite the current technical limitations for producing graphene structures with perfect edge terminations, we expect that future advances in lithography will allow the patterning of devices with the required precision for the development of a device based on the structure described here.

\acknowledgements Discussions with M. Petrovic and D. R. da Costa are gratefully acknowledged. This work was financially supported by CNPq, under the Science Without Borders and PRONEX/FUNCAP grants, CAPES and Lemann Foundation.

\end{document}